\documentclass[12pt]{iopart}

\usepackage{iopams}
\usepackage{amssymb}
\usepackage{graphicx}
\eqnobysec

\begin{document}

\newcommand{\sign}{\mathrm{sign}}

\title[Correlations and screening of
topological charges in gaussian random fields]{Correlations and screening of
topological charges in gaussian random fields}

\author{M R Dennis}

\address{H H Wills Physics Laboratory, Tyndall Avenue, Bristol BS8 1TL, UK}

\begin{abstract}
2-point topological charge correlation functions of several types of geometric singularity in gaussian random fields are calculated explicitly, using a general scheme:  
zeros of $n$-dimensional random vectors, signed by the sign of their jacobian determinant; critical points (gradient zeros) of real scalars in two dimensions signed by the hessian; and umbilic points of real scalars in two dimensions, signed by their index. The functions in each case depend on the underlying spatial correlation function of the field. These topological charge correlation functions are found to obey the first Stillinger-Lovett sum rule for ionic fluids.
\end{abstract}

\section{Introduction}\label{sec:int}

Although a spatially extended field may be smooth, and contain no
infinities, it may nonetheless have point singularities associated
with its topology. For example, a 2-dimensional landscape,
specified by a real function, has \emph{critical points}
(stationary points), where the gradient of the field vanishes                                                                                                                                                                                                                                                                                                                                                                   and
the gradient direction cannot be defined; in the neighbourhood of
such a point, the gradient direction changes arbitrarily quickly
through all of its values. Such points are characterised by a
\emph{topological charge}, a signed integer which is determined by
the local geometry of the singularity; in the case of critical
points, the number is the (signed) number of rotations of the
gradient vector around the critical point; it is $+1$ for maxima and minima, and
$-1$ for saddles. The topological charge is also called the
\emph{Poincar\'e index} (or \emph{Hopf index}) (Milnor 1965), and is defined as a signed integer at the point zeros of $n$-dimensional real vector fields in $n$ dimensions.  
Although the number of zeros may change
as a field evolves, the total topological charge is constant; it is a topological invariant.

Here, I discuss the statistical properties of various zeros/singularities, in fields which are specified by gaussian random
functions. In this case, only structurally stable zeros of charge $\pm 1$ occur. Specifically, I shall discuss how densities and two
point charge correlation functions of the distributions of signed
singularities may be calculated under a rather general scheme, and
then use this scheme to calculate the topological charge correlation functions
of three types of singularity in isotropic random fields: zeros of
$n$-dimensional vectors in  $n$-dimensional space (rederiving a
result originally due to Halperin (1981));
critical points of random scalar fields in two dimensions; and umbilic
points of random scalar fields in two dimensions. The charge correlation
functions for each of these types of topological singularity are
found to be different in each case, and dependent on the
underlying correlation function of the gaussian field. Each
function is found to satisfy a screening relation 
associated with ionic liquids. 

Topological zeros are very important in many areas of physics and
mathematics: in addition to critical points (gradient zeros) which
have obvious importance, the zeros of 2-dimensional complex scalar fields (phase singularities, wave dislocations, vortices, realised as 2-dimensional vectors) are
also of great interest, especially where the field is a quantum wavefunction
or an optical field (e.g. Nye and Berry 1974, Berry and Dennis 2000, Dennis 2001b), or an order parameter (Mermin 1979). Only point zeros are considered here - the
dimensionality $n$ of the vector field, whose zeros provide the
singularity, is assumed equal to the dimensionality of the space.

The first systematic study of the statistical geometry of random real scalar 
fields in two dimensions was by Longuet-Higgins (1957a,b, 1958), who generalised
one-dimensional methods of Rice (1954) to calculate, amongst other things, the densities of critical points and probability density function of the gaussian curvature of the function. Halperin (1981) derived the $n$-dimensional vector correlation function, whose proof was supplied by Liu and Mazenko (1992), and has recently been recast in the language of riemannian geometry by Foltin (2003a). Various correlation functions in two dimensions, including those for phase singularities and critical points, were investigated numerically by Freund and Wilkinson (1998). Planar phase singularity correlations (including density correlations) were investigated by Berry and Dennis (2000). 

The topological singularities of interest are the zeros of an
$n$-dimensional vector field $\bi{v} = (v_1,\dots, v_n).$ The
field is a smooth function defined on an $n$-dimensional euclidean
space, with points labelled by the vector $\bi{r} = (r_1,\dots,
r_n).$ In random fields, the only statistically significant zeros
are those of first order, whose jacobian determinant
\begin{equation}
   \mathcal{J} = \det \partial_i v_j,
   \label{eq:jdef}
\end{equation}
is nonzero (where $\partial_i \equiv \bullet_{,i} \equiv
\partial/\partial r_i$). The topological charge of such zeros is
given by $\sign \mathcal{J}.$

When the field is random, the density $d(\bi{r})$ of zeros at a
position $\bi{r}$ is
\begin{equation}
   d(\bi{r}) = \langle \delta^n(\bi{v}(\bi{r})) |\mathcal{J}(\bi{v}(\bi{r}))| \rangle,
   \label{eq:densdef}
\end{equation}
where $\langle \bullet \rangle$ denotes averaging over the
statistical ensemble, and the zeros are identified by the
$n$-dimensional Dirac $\delta$-function. The modulus of the
jacobian is included to ensure that each zero has the correct
statistical weight, so $d(\bi{r})$ has the units of density. This
expression confirms that degenerate zeros are so rare they have no
statistical significance. The average topological charge
$q(\bi{r})$ at $\bi{r}$ is expressed as the density
(\ref{eq:densdef}), but each singularity is weighted by its
charge, that is, the sign of the jacobian. The charge correlation functions $g$ with which this paper is
concerned are the generalisation of the average
charge density to that at two points $\bi{r}_A, \bi{r}_B,$ normalised by
the density:
\begin{equation}
   g(\bi{r}_A,\bi{r}_B) = \frac{1}{d(\bi{r}_A) d(\bi{r}_B)} \langle \delta^n(\bi{v}(\bi{r}_A))
    \delta^n(\bi{v}(\bi{r}_B))\mathcal{J}(\bi{v}(\bi{r}_A))
   \mathcal{J}(\bi{v}(\bi{r}_B)) \rangle.
   \label{eq:gdef}
\end{equation}

In the following section, the scheme for calculating topological charge correlation functions (\ref{eq:gdef}) in general gaussian random fields is explained. Section \ref{sec:fields} then provides background to gaussian fields which are statistically stationary and isotropic. The scheme is then applied to zeros of $n$-dimensional vector fields (section \ref{sec:vec}), critical points of 2-dimensional scalar fields (section \ref{sec:crit}), and umbilic points of the same fields (section \ref{sec:umbilic}). The phenomenon of topological charge screening in these three cases is discussed in section \ref{sec:screening}.

As this paper was being completed, I became aware of the work of Foltin (2003b), who performs similar calculations for critical points by a different, possibly simpler method.

\section{Gaussian random functions and scheme for calculating charge correlation functions}\label{sec:scheme}

An ensemble of scalar functions $f$ in $n$-dimensional space is
said to be a {\itshape gaussian random function} if the
probability distribution of the function at each point of space is
given by a gaussian distribution, so the function at each point
defines a gaussian random variable (Adler 1981). The
only restrictions on gaussian random functions made in this
section are that they be centred (the average $\langle f \rangle =
0$) and that first derivatives exist.

Our starting point is the well-known expression for the probability
density function for a set of $N$ independent gaussian
random functions $\bi{u} = (u_1,\dots,u_N),$
\begin{equation}
   P(\bi{u}) = \frac{\exp(-\bi{u} \cdot \bSigma^{-1} \cdot \bi{u}/2)}{(2\pi)^{N/2} \sqrt{\det
   \bSigma}},
   \label{eq:gausspdf1}
\end{equation}
where $\bSigma$ is the {\itshape correlation matrix}, with components defined
\begin{equation}
   \Sigma_{ij} = \langle u_i u_j \rangle.
   \label{eq:cormatdef}
\end{equation}

It is now possible to present the general scheme for calculating
2-point topological charge correlation functions of zeros of
gaussian random functions, expressed in (\ref{eq:gdef}). The
topological charges considered are all zeros of a real vector
gaussian random function $\bi{v} = (v_1,\dots, v_n),$ of dimension
$n,$ and the correlation function is the average of the product of
the local density at points labelled $A,B.$ Quantities evaluated
at these places are denoted with the appropriate subscript $A$ or
$B.$ The variables that appear in the average (\ref{eq:gdef}) are
therefore the components of $\bi{v}_A, \bi{v}_B,$ and their
derivatives $\partial \bi{v}_A, \partial \bi{v}_B$ that appear in
the jacobians $\mathcal{J}(\bi{v}_A) \equiv \mathcal{J}_A,
\mathcal{J}(\bi{v}_B) \equiv \mathcal{J}_B.$ There are a total of
$m$ different derivatives $\partial_i v_j$ appearing in each
jacobian, and in general $m \leq n^2.$ For example, in the case of
critical points, $n = 2$, and $m = 3$ (the terms appearing in the
jacobian are $\partial_{11} f, \partial_{22} f,
\partial_{12} f = \partial_{21} f$). The calculation therefore
involves an average of the $N= 2(m+n)$-dimensional gaussian random
vector
\begin{equation}
\fl \bi{u} = \left( (\partial \bi{v}_A)_1, \dots, (\partial
   \bi{v}_A)_m,(\partial \bi{v}_B)_1,
   \dots, (\partial \bi{v}_B)_m, v_{A1},\dots, v_{An}, v_{B1},\dots, v_{Bn}\right).
   \label{eq:bigudef}
\end{equation}
The correlation matrix $\bSigma$ for this $\bi{u}$ is defined as in
(\ref{eq:cormatdef}), and averages are evaluated according to the
probability density function (\ref{eq:gausspdf1}).

The correlation function in (\ref{eq:gdef}) is therefore
\begin{eqnarray}
\fl  g_{AB} = \frac{\langle \delta^n(\bi{v}_A) \delta^n(\bi{v}_B)
\mathcal{J}_A \mathcal{J}_B
   \rangle}{d(\bi{r}_A) d(\bi{r}_B)} \nonumber \\
\lo= \frac{1}{d(\bi{r}_A) d(\bi{r}_B) (2\pi)^{n+m} \sqrt{\det \bSigma}} \nonumber \\
\times \int \rmd^{2(n+m)} \bi{u} \,
   \delta^n(\bi{v}_A) \delta^n(\bi{v}_B) \mathcal{J}_A \mathcal{J}_B \exp(-\bi{u} \cdot
   \bSigma^{-1} \cdot
   \bi{u}/2).
   \label{eq:gab1}
\end{eqnarray}
where $d(\bi{r}_A), d(\bi{r}_B)$ are the appropriate densities of zeros (\ref{eq:densdef}). The jacobians
$\mathcal{J}_A, \mathcal{J}_B$ can be quite complicated; each is a
multilinear function, involving a sum of $n$-fold products of
$u_1, \dots, u_m$ (for $A$) or $u_{m+1}, \dots, u_{2m}$ (for $B$).

The $2n\times 2n$-dimensional lower right submatrix of $\bSigma$
(i.e. the averages dependent only on $\bi{v}_A, \bi{v}_B$) is
denoted by $\mathbf{K};$ the complementary $2m\times
2m$-dimensional upper left submatrix of $\bSigma^{-1}$ is denoted
by $\bXi^{-1};$ that is,
\begin{equation}
   \bSigma = \left( \begin{array}{c|c} \bullet  & \bullet \\
   \hline \bullet & \mathbf{K}
   \end{array} \right), \qquad
   \bSigma^{-1} = \left( \begin{array}{c|c} \bXi^{-1}  & \bullet \\
   \hline \bullet & \bullet
   \end{array} \right),
   \label{eq:sigkxidef}
\end{equation}
where the other blocks marked $\bullet$ are not denoted by special
symbols. In \ref{sec:jacobi}, Jacobi's determinant
theorem is used to show that
\begin{equation}
   \det\bSigma = \det\mathbf{K} \, \det \bXi.
   \label{eq:jacobi}
\end{equation}

The integrals in $v_{A1}, \dots, v_{Bn},$ only involving
$\delta$-functions, are performed, leaving the $2m$ derivative terms
$\partial\bi{v}$ to integrate; let $\bi{u}' =
((\partial\bi{v}_A)_1,\dots, (\partial\bi{v}_B)_m).$ Then
\begin{eqnarray}
   \fl   g_{AB} =  \frac{1}{d(\bi{r}_A) d(\bi{r}_B) (2\pi)^{n+m} \sqrt{\det \bSigma}} \int \rmd^{2m}\bi{u}' \,
   \mathcal{J}_A \mathcal{J}_B \exp(-\bi{u}'\cdot \bXi^{-1} \cdot \bi{u}'/2) \nonumber \\
   \lo=   \frac{1}{d(\bi{r}_A) d(\bi{r}_B) (2\pi)^{n+m} \sqrt{\det \bSigma}} \nonumber \\
\times \int \rmd^{2m}\bi{u}' \,
   \mathcal{J}_A \mathcal{J}_B  \frac{\sqrt{\det \bXi}}{(2\pi)^m} \int
   \rmd^{2m}\bi{t} \, \exp(\rmi \bi{t} \cdot \bi{u}'
   - \bi{t}\cdot \bXi\cdot \bi{t}/2), \label{eq:gab2}
\end{eqnarray}
upon Fourier transforming the gaussian in $\bi{u}'$ with the
$2m$-dimensional Fourier vector variable $\bi{t}.$ The jacobian
terms $\mathcal{J}_A$ (depending on $u'_1, \dots, u'_m$) and
$\mathcal{J}_B$ (depending on $u'_{m+1},\dots,u'_{2m}$) may be
replaced with partial derivative terms in $t_j,$
\begin{equation}
   u'_j \to\rmi \nabla_{t_j} = \rmi \nabla_j,
   \label{eq:ftu}
\end{equation}
where $\nabla$ is used to denote partial derivatives in
$2m$-dimensional $\bi{t}$-space. The jacobians, with this replacement, have now become
differential operators, denoted $\mathcal{J}_{\nabla A},
\mathcal{J}_{\nabla B}.$
Therefore
\begin{eqnarray}
   \fl   g_{AB} = \frac{1}{d(\bi{r}_A) d(\bi{r}_B) (2\pi)^{n+2m}} \sqrt{\frac{\det \bXi}{\det \bSigma}} \nonumber \\ \times
   \int\rmd^{2m}\bi{t} \, \exp(-\bi{t}\cdot \bXi \cdot \bi{t}/2) \mathcal{J}_{\nabla A}
   \mathcal{J}_{\nabla B} \int \rmd^{2m}\bi{u}' \, \exp(\rmi \bi{t} \cdot \bi{u}') \nonumber \\
   \lo=   \frac{1}{d(\bi{r}_A) d(\bi{r}_B) (2\pi)^{n+2m}} \frac{(2\pi)^{2m}}{\sqrt{\det \mathbf{K}}} \nonumber \\
   \times \int \rmd^{2m}\bi{t} \, \exp(-\bi{t}\cdot \bXi \cdot \bi{t}/2) \mathcal{J}_{\nabla A}
   \mathcal{J}_{\nabla B} \delta^{2m}(\bi{t}),
   \label{eq:gab3}
\end{eqnarray}
where, in the second line, the integral over $\bi{u}'$ is realised
as the Fourier transform of the $\delta$-function of $\bi{t},$ and
(\ref{eq:jacobi}) has been used in the prefactor. The expression
is then integrated by parts (so the $\mathcal{J}_{\nabla}$
operators act on the exponential term rather than the
$\delta$-function), and then integrated in $\bi{t}.$ The final
expression is
\begin{equation}
   g_{AB} = \frac{1}{d(\bi{r}_A) d(\bi{r}_B)(2\pi)^n} \frac{1}{\sqrt{\det \mathbf{K}}} D,
   \label{eq:gabfin}
\end{equation}
where
\begin{equation}
   D =  \left[ \mathcal{J}_{\nabla A}\mathcal{J}_{\nabla B} \exp(-\bi{t}\cdot
   \bXi \cdot \bi{t}/2)
  \right]_{\bi{t} = \mathbf{0}}.
   \label{eq:ddef}
\end{equation}
The charge correlation function therefore only depends on $d, \det
\mathbf{K}$ and components of the inverse reduced inverse
correlation matrix $\bXi.$

This final step, evaluating $D$ in (\ref{eq:ddef}) is the most
complicated part of the calculation, and depends on the precise
form of the jacobian determinant $\mathcal{J}.$ Each summand in
the operator $\mathcal{J}_{\nabla A} \mathcal{J}_{\nabla B}$ is a
$2n$-fold derivative over some index set $\{\tau\} = \{\tau_1,
\dots, \tau_{2n}\},$ and is comprised of $n$ terms from $\nabla_1,
\dots, \nabla_m$ (for $A$), and $n$ terms from $\nabla_{m+1},
\dots, \nabla_{2m}$ (for $B$), either set possibly including
repetitions. The result of this particular
operation is
\begin{equation}
   \left[\left(\prod_{j=1}^{2n}\rmi  \nabla_{\tau_j}\right) \exp(-\bi{t} \cdot
   \bXi \cdot \bi{t}/2) \right]_{\bi{t} = \mathbf{0}} =  \sum^{{\rm all\, pairings}}_{p}
\prod_{(\tau_{\lambda},\tau_{\lambda'}) \in p} \Xi_{\tau_{\lambda}
\tau_{\lambda'}},
   \label{eq:dexp}
\end{equation}
where the sum on the right hand side is over all pairings $p =
(\tau_{\alpha}, \tau_{\alpha'}) \dots (\tau_{\mu}, \tau_{\mu'})$
of indices in $\{\tau\};$ there are $(2n-1)!!$ such pairings, each
involving $n$ sets of pairs. The $n$-fold product on the right
hand side is over all components of $\bXi$ with indices given by
the appropriate pair. The $\Xi_{ij}$ themselves are found by
further application of Jacobi's determinant theorem in 
\ref{sec:jacobi}, and are expressed in terms of minors of
$\det\bSigma$ in (\ref{eq:xiij}).

The correlation functions calculated explicitly in this paper are
not too complicated, either because $n$ is small (only two-dimensional fields are considered in sections \ref{sec:crit},
\ref{sec:umbilic}), or $\bXi$ is sparse, as in the case of random
vectors (section \ref{sec:vec}).

The density $d$ of zeros appears in (\ref{eq:gabfin}); in general,
this may be difficult to calculate, due to the modulus sign in
(\ref{eq:densdef}). For random $n$-dimensional vector fields, the main part of the density calculation is in \ref{sec:dveccalc}. For critical and umbilic points, these densities were calculated by Longuet-Higgins (1957a,b) and Berry and Hannay (1977).

\section{Isotropic random fields}\label{sec:fields}

The topological charge correlation function in (\ref{eq:gabfin})
is extremely general, applying to any centred differentiable
gaussian random vector field. In this section, and for the
remainder of the article, attention will be restricted to
stationary isotropic random fields. For these fields, all averages
are (statistically) translation and rotation invariant. They are
conveniently given by a Fourier representation
\begin{equation}
   f(\bi{r}) = \sum_{\bi{k}} a(k) \cos(\bi{k} \cdot \bi{r} + \phi_{\bi{k}}),
   \label{eq:ffourier}
\end{equation}
where $\bi{k}$ are now the Fourier variable vectors (wavevectors).
The components $v_i$ of random vectors are specified by
independent identically distributed realisations of
(\ref{eq:ffourier}). The amplitude $a(k)$ only depends on the
magnitude $|\bi{k}| = k,s$ and the phase
$\phi_{\bi{k}}$ is uniformly random - each ensemble member is
therefore labelled by the choice of $\phi_{\bi{k}}$ for each
$\bi{k}.$ It also may represent the spatial part of a real linear
homogeneous nondispersive wavefield, for which the representation
(\ref{eq:ffourier}) is particularly evocative. The infinite
$\bi{k}$ set is assumed sufficiently dense that they may be
represented as an integral, and
\begin{equation}
   \sum_{\bi{k}} a^2(k) \bullet \approx \int \rmd^n \bi{k} \, \Pi(k) \bullet,
   \label{eq:pidef}
\end{equation}
where $\Pi(k)$ is the power spectrum of the field; by the
Wiener-Khinchine theorem (Feller 1950), $\Pi(k)$ is
the $n$-dimensional Fourier transform of the \emph{field correlation
function} $C(r),$ where $r = |\bi{r}_A - \bi{r}_B|,$ and
\begin{equation}
   C(r) = \langle f_A f_B \rangle = \langle f(\bi{r}_A) f(\bi{r}_B) \rangle,
   \label{eq:cdef}
\end{equation}
normalised such that $C(0) = \langle f^2 \rangle = 1.$ The only
condition on $C$ is that it is symmetric and has positive Fourier
transform.

Averages of derivatives of $f$ may be represented as moments of
$\Pi$ (as by Longuet-Higgins 1957a,b, Berry and Hannay 1977, Berry and Dennis 2000), or
equivalently derivatives of $C$ as is done here. Coordinates are chosen where
\begin{equation}
   \bi{r} = \bi{r}_{B} - \bi{r}_A, \quad r_1 = r, \quad r_j = 0, \; j = 2, \dots, n.
   \label{eq:rfix}
\end{equation}
Since the fields are isotropic, the results are not affected by
this choice. The correlations computed in sections \ref{sec:crit},
\ref{sec:umbilic} are in two dimensions; in this case, direction 1
is denoted by $x,$ direction 2 by $y.$

Representing derivatives by subscripts, the
correlations of first derivatives of $f$ are found to be (Berry and Dennis 2000)
\begin{eqnarray}
\fl   E  & \equiv \langle f_{A,1} f_B \rangle = - \langle f_A f_{B,1} \rangle
   & = -\left[\partial_1 C\right]_{r_1 = r, r_2,\dots = 0} = -C' \nonumber \\
\fl   F  & \equiv \langle f_{A,1} f_{B,1} \rangle = -\langle f_{A,11} f_B \rangle
   = -\langle f_A f_{B,11} \rangle & = -\left[\partial_1^2 C\right]_{r_1 = r, r_2,\dots = 0}= -C'', \nonumber \\
\fl   H  & \equiv \langle f_{A,j} f_{B,j} \rangle = -\langle f_{A,jj} f_B \rangle
   = -\langle f_A f_{B,jj} \rangle & = -\left[\partial_2^2 C\right]_{r_1 = r, r_2,\dots = 0} = -C'/r, \quad j\neq
   1.
   \label{eq:1stcorrs}
\end{eqnarray}
Averages involving $f$ and its first derivatives other than those
in (\ref{eq:cdef}), (\ref{eq:1stcorrs}) are zero. 
The averages are equal to appropriate derivatives of $C(\bi{r}),$ and then setting $r_1 = r, r_2, \dots = 0$ (as in (\ref{eq:rfix})), and derivatives in $r_j$ of odd order vanish. The functions in (\ref{eq:1stcorrs}),
when $\bi{r} = \mathbf{0}$ (denoted by subscript 0), are
\begin{equation}
   E_0 = 0, \qquad F_0 = H_0 = -C''(0) = -C''_0.
   \label{eq:1stcorr0}
\end{equation}
It is easily verified that $-C_0''>0,$ since $C$ has a positive
Fourier transform. The correlation matrices (\ref{eq:cormatcrit}),
(\ref{eq:cormatumb}) required to calculate the topological charge
correlation functions of critical points and umbilic points,
involve higher derivatives of $C,$ given in (\ref{eq:2ndcorrs}),
(\ref{eq:3rdcorrs}).

As the separation between $A$ and $B$ increases, the correlation
between $f_A$ and $f_B$ decreases and $C\to 0.$ This decay is
slowest in the case where all the $\bi{k}$ in (\ref{eq:ffourier})
have the same length $k_0,$ (set to 1 for convenience) and the power spectrum $\Pi(k) =
\delta(k-1).$ For any $n>2,$ the corresponding correlation
function is a Bessel function times a factor of $r,$ with
oscillatory decay that falls off like $r^{-(n-1)/2}.$ Of
particular interest is the $n=2$ case, for which $C(r) = J_0(r).$
The spectrum in this case was called the ring spectrum by Longuet-Higgins (1957a,b) and Berry and Dennis (2000), and is conjectured to
model the high eigenfunctions in quantum chaotic billiards
(Berry 1978, 2002).

\section{Correlations of zeros of isotropic vector fields}\label{sec:vec}

In the present section, we shall consider $n$-dimensional gaussian
random vector fields $\bi{v} = (v_1,\dots,v_n)$ in $n$ dimensions
whose cartesian components are independent and identically
distributed gaussian random fields (the derivatives of the
components are also assumed completely independent). The jacobian
$\mathcal{J},$ whose sign determines the topological charges of
the zeros, is the determinant of the matrix of first derivatives
(\ref{eq:jdef}).

We begin by calculating the density (\ref{eq:densdef}) of zeros of
random vectors. This was calculated for $n = 1,2,3$ by Halperin (1981) and Liu and Mazenko (1992), (the $n = 1$ case was
previously found by Rice (1954), and $n = 2$ by Berry (1978)). For general $n,$ the density $d_n$
(\ref{eq:densdef}) is expressed as a probability integral with
density function (\ref{eq:gausspdf1}) and correlations given by
(\ref{eq:1stcorrs}),(\ref{eq:1stcorr0}). Therefore
\begin{eqnarray}
\fl d_n = \langle \delta^n(\bi{v}_i) |\det v_{i,j} | \rangle\nonumber \\
\fl \qquad= \frac{1}{(2\pi)^{n(n+1)/2} F_0^{n^2/2}} \int \rmd^n
v_i
   \rmd^{n^2} v_{i,j} \, \delta^n(\bi{v}) |\det v_{i,j} |
   \exp\left(-\frac{1}{2}\sum_{i=1}^n v_i^2 -\frac{1}{2}
   \sum_{i,j=1}^n v_{i,j}^2\right) \nonumber \\
   \lo= \frac{F_0^{n/2}}{(2\pi)^{n(n+1)/2}} \int\rmd^{n^2} v_{i,j} \,
   |\det v_{i,j} | \exp\left(-\frac{1}{2}\sum_{i,j=1}^n v_{i,j}^2\right),
   \label{eq:dveccalc}
\end{eqnarray}
where in the final line the $\delta$-functions in the
$v_i$ were integrated, and the $v_{i,j}$ were rescaled (each by
$\sqrt{F_0}$) to be dimensionless. The remaining integral is
solved in \ref{sec:dveccalc}, giving the density of zeros
in $n$ dimensions
\begin{equation}
   d_n = (-C_0'')^{n/2} \frac{(\frac{n-1}{2})!}{\pi^{(n+1)/2}} =
   (-C_0'')^{n/2} \frac{(n-1)! \sigma_{n-1}}{(2\pi)^n}.
   \label{eq:dvec}
\end{equation}
In this expression, $\sigma_{n-1}$ is the surface area of the unit
$(n-1)$-sphere in $n$ dimensions, given by (\ref{eq:spherearea}).
As is common in such problems in statistical geometry, the result
is a spectral quantity ($(-C_0'')^{n/2}$) multiplied by a geometric
factor.

The scheme of the previous section may now
be applied to calculate the topological charge correlation
function for zeros in these gaussian random vector fields.
Implementation of the scheme is facilitated by the fact that the
components $v_i$ are completely independent, and $m = n^2.$ The
submatrix $\mathbf{K}_n$ of the correlation matrix $\bSigma_n$
only depends on the correlations of the components of the vectors
$\bi{v}_A, \bi{v}_B;$ from section \ref{sec:fields}, the only
correlations that do not vanish are $\langle v_{Ai}^2 \rangle =
\langle v_{Bi}^2 \rangle = 1$ and $\langle v_{Ai} v_{Bi} \rangle =
C$ for $i = 1,\dots, n.$ It is easy to see that
\begin{equation}
   \det \mathbf{K}_n = (1-C^2)^n.
   \label{eq:vecdetk}
\end{equation}
From (\ref{eq:dexp}), the other necessary ingredient of the correlation
function scheme is the components of the matrix $\bXi,$ defined in
(\ref{eq:sigkxidef}). The elements of $\bXi$ are labelled by the
multiindex of the components $v_{Ai,j}, v_{Bk,l};$ using the
correlations (\ref{eq:1stcorrs}), (\ref{eq:1stcorr0}) and the
arguments of \ref{sec:jacobi}, particularly equation
(\ref{eq:xiij}), the only nonvanishing elements are
\begin{eqnarray}
   \Xi_{(Ai,1)(Ai,1)} = \Xi_{(Bi,1)(Bi,1)} = F_0 - E^2/(1-C^2)
   \nonumber\\
   \Xi_{(Ai,1)(Bi,1)} = \Xi_{(Bi,1)(Ai,1)} = F_0 - E^2/(1-C^2)
   \nonumber \\
   \Xi_{(Ai,j)(Ai,j)} = \Xi_{(Bi,j)(Bi,j)} = F_0 \nonumber \\
   \Xi_{(Ai,j)(Bi,j)} = \Xi_{(Bi,j)(Ai,j)} = H \qquad
   \mathrm{for\,} i=1,\dots,n, \, \, j=2,\dots, n.
   \label{eq:xivec}
\end{eqnarray}

The problem remains to use these components and (\ref{eq:dexp}) to
evaluate (\ref{eq:ddef}). By (\ref{eq:jdef}), (\ref{eq:ftu}),
\begin{equation}
   \mathcal{J}_{\nabla A} = \det \nabla_{Ai,j},
   \label{eq:jvecdet}
\end{equation}
(similarly for $\mathcal{J}_{\nabla B}$). Each of the $n!$
summands in this determinant is an $n$-fold product $\sign \sigma
\prod_i \nabla_{Ai,\sigma(i)},$ where $\sigma$ is a permutation of
$1,\dots,n.$ Therefore
\begin{equation}
   \mathcal{J}_{\nabla A} \mathcal{J}_{\nabla B} = \sum_{\sigma,
   \sigma'}^{\mathrm{permutations}} \sign \sigma \, \sign \sigma' (-1)^n \prod_{i,j=1}^n
   \nabla_{Ai,\sigma(i)} \nabla_{Bj,\sigma(i)}.
   \label{eq:vecjacop}
\end{equation}
From (\ref{eq:dexp}), the result of one of these summands acting
on $\exp(-\bi{t}\cdot\bXi\bi{t}/2)$ and setting $\bi{t} =
\mathbf{0}$ is nonzero if there is a pairing of these multiindices
where the corresponding elements of $\bXi$ are nonzero. From
(\ref{eq:xivec}), this is only the case when the permutations
$\sigma, \sigma'$ are the same. Thus, from (\ref{eq:dexp}) and
(\ref{eq:xivec}),
\begin{eqnarray}
   D_n & = & \sum_{\sigma}^{\mathrm{permutations}}\prod_{i=1}^n
   \Xi_{(Ai,\sigma(i)) (Bi,\sigma(i))} \nonumber \\
   & = &  n! \prod_{i=1}^n \Xi_{(Ai,i) (Bi,i)} \nonumber \\
   & = &  n! (F - C E^2/(1-C^2)) H^{n-1}.
   \label{eq:dn}
\end{eqnarray}
This, together with (\ref{eq:vecdetk}), can now be put into
(\ref{eq:gabfin}) to give
\begin{eqnarray}
   g_n(r) & = & \frac{n!}{2 \pi d_n^2} \frac{(F(1 - C^2) - C E^2)}{(1-C^2)^{3/2}}
   \left( \frac{H}{2\pi \sqrt{1-C^2}} \right)^{n-1} \nonumber \\
   & = & - \frac{n!}{2\pi d_n^2} \frac{(C'' (1-C^2) + C C'^2)}{(1-C^2)^{3/2}}
   \left( \frac{-C'}{2\pi r \sqrt{1-C^2}} \right)^{n-1}\nonumber
   \\
   & = & \frac{(n-1)!}{(2\pi)^n d_n^2 r^{n-1}} \frac{\rmd h_n(r)}{\rmd r},
   \label{eq:vecgab}
\end{eqnarray}
where the function $h_n(r)$ is defined
\begin{equation}
   h_n(r) = \left( \frac{-C'}{\sqrt{1 - C^2}}\right)^n.
   \label{eq:hndef}
\end{equation}
This decays to 0 as $r\to\infty,$ and when $r = 0,$
\begin{equation}
   h_n(0) = (-C''_0)^{n/2}
   \label{eq:hn0}
\end{equation}
(the quantity $C''_0 = C''(0)$ is always negative, since $C$ is
the Fourier transform of the positive power spectrum $\Pi$).

\begin{figure}
\begin{center}
\includegraphics*[width=12cm]{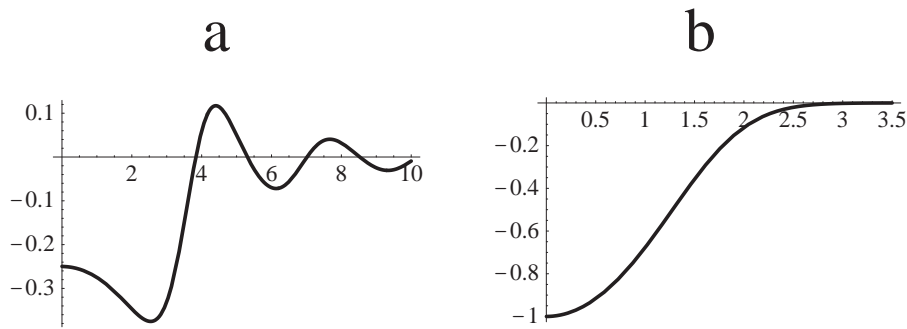}
\end{center}
    \caption{The 2-dimensional vector zero correlation function $g_2(r),$ plotted for two
    choices of $C(r):$ (a) $C(r) = J_0(r);$ (b) $C(r) = \exp(-r^2/2).$}
    \label{fig:g2}
\end{figure}

The topological charge correlation function of zeros of
$n$-dimensional gaussian random vector fields $g(r)$ was first
calculated by Halperin (1981), in a form
equivalent to (\ref{eq:vecgab}), but without proof. This function
was also derived by different means by Liu and Mazenko (1992), and in the $n = 1$ case by Rice (1954), and the $n = 2$ case by Berry and Dennis (2000) and Foltin (2003a). $g(r)$ is plotted in figure \ref{fig:g2} for
two choices of the field correlation function $C$ for $n = 2.$ When $C(r) = J_0(r),$ $g_2$ is oscillatory; when $C(r) = \exp(-r^2/2),$ it increases monotonically to zero as $r\to\infty.$

\section{Critical points in two dimensions}\label{sec:crit}

In this section, the scheme of section \ref{sec:scheme} is used to
calculate the topological charge correlation function of critical
points of isotropic gaussian random functions in the plane, that
is the Poincar{\'e} index correlation function of isotropic random
surfaces. 

The gaussian random function examined shall be written $f =
f(\bi{r}) = f(x,y)$ (where $x = r_1, y = r_2$), which will be
assumed stationary and isotropic, so the expressions in section
\ref{sec:fields} may be used. In particular, $f$ and its first
derivatives have the correlations (\ref{eq:1stcorrs}),
(\ref{eq:1stcorr0}), as well as further correlations involving
second derivatives, described below. As in the previous section,
for convenience in calculations, the two points $A$ and $B$ are
separated only in the $x$ coordinate.

At a critical point, the gradient $\nabla f = (f_{,x},  f_{,y})$
is zero. The critical point jacobian $\mathcal{J}_{\mathrm{c}}$
whose sign defines the topological charge (Poincar{\'e} index) is
the hessian determinant
\begin{equation}
   \mathcal{J}_{\mathrm{c}} = \det f_{,ij} =  f_{,xx}  f_{,yy} -
   f_{,xy}^2;
   \label{eq:hessdef}
\end{equation}
this is the gaussian curvature of the surface. Unlike a general
2-dimensional random vector field (as in the previous section),
the gradient field $\nabla f$ is irrotational, which gives rise to
relationships and correlations between the derivatives of the
components (e.g. $f_{,xy} = f_{,yx}$, whereas before, $v_{1,2}
\neq  v_{2,1}$ in general). This makes the explicit computation of
the topological charge correlation function more difficult than in
the previous section, even in the 2-dimensional case that is
considered here; the scheme of section \ref{sec:scheme} applies to
gradient zeros in fields of any dimension.

The statistical properties of critical points of gaussian random
fields in 2 dimensions were considered by Longuet-Higgins (1957a,b); he found that the density
$d_{\mathrm{c}}$ of critical points is (Longuet-Higgins 1957b equations (71), (78)):
\begin{equation}
   d_{\mathrm{c}} = \left| \frac{2 C^{(4)}_0}{3\pi\sqrt{3} C''_0} \right|,
   \label{eq:critdens}
\end{equation}
($C^{(4)}_0$ denotes the fourth derivative of $C,$ evaluated at $r=0$).  The density of saddles equals the density of extrema
(maxima and minima), and the density of maxima equals the density
of minima. The probability density function of the gaussian
curvature $\mathcal J$, despite its asymmetry
(Longuet-Higgins 1958 equation (7.14), Dennis 2002 equation (57)), has zero first
moment, implying the average topological charge $\langle
\delta^2(\nabla f) \mathcal{J}_{\mathrm{c}} \rangle$ is zero, as
expected.

The topological charge correlation function is again calculated
using the scheme of section \ref{sec:scheme}, particularly
(\ref{eq:gabfin}). Therefore, the vector $\bi{u}$
(\ref{eq:bigudef}) of gaussian random variables, in a convenient
ordering, is
\begin{equation}
   \bi{u}_{\mathrm{c}} = (f_{A,xx}, f_{A,yy}, f_{B,xx}, f_{B,yy}, f_{A,xy},f_{B,xy},
   f_{A,x}, f_{B,x}, f_{A,y}, f_{B,y})
   \label{eq:bigucrit}
\end{equation}
with correlation matrix (c.f. (\ref{eq:cormatdef}))
\begin{equation}
   \bSigma_{\mathrm{c}} = \left( \begin{array}{cccccccccc}
M_0 & L_0 &   M &   L &   0 &   0 &   0 &   G &   0 &   0 \\
L_0 & M_0 &   L &   N &   0 &   0 &   0 &   I &   0 &   0 \\
  M &   L & M_0 & L_0 &   0 &   0 &  -G &   0 &   0 &   0 \\
  L &   N & L_0 & M_0 &   0 &   0 &  -I &   0 &   0 &   0 \\
  0 &   0 &   0 &   0 & L_0 &   L &   0 &   0 &   0 &   I \\
  0 &   0 &   0 &   0 &   L & L_0 &   0 &   0 &  -I &   0 \\
  0 &   0 &  -G &  -I &   0 &   0 & F_0 &   F &   0 &   0 \\
  G &   I &   0 &   0 &   0 &   0 &   F & F_0 &   0 &   0 \\
  0 &   0 &   0 &   0 &   0 &  -I &   0 &   0 & F_0 &   H \\
  0 &   0 &   0 &   0 &   I &   0 &   0 &   0 &   H & F_0 \\
\end{array} \right),
   \label{eq:cormatcrit}
\end{equation}
where the correlations between elements of $\bi{u}_{\mathrm{c}}$
are computed to be
\begin{eqnarray}
   G & \equiv \left[\partial_x^3 C\right]_{x=r,y=0} & = C^{(3)} \nonumber \\
   I & \equiv \left[\partial_x^2 \partial_y C\right]_{x=r,y=0} & = (r C'' - C')/r^2 \nonumber \\
   L & \equiv \left[\partial_x^2 \partial_y^2 C\right]_{x=r,y=0} & = (r^2 C^{(3)} - 2 r C'' + 2 C')/r^3 \nonumber \\
   M & \equiv \left[\partial_x^4 C\right]_{x=r,y=0} & = C^{(4)}\nonumber \\
   N & \equiv \left[\partial_y^4 C\right]_{x=r,y=0} & = 3(r C'' - C')/r^3 \nonumber \\
   G_0 & = I_0 = 0, & M_0 = N_0 = 3L_0 = C^{(4)}_0.
   \label{eq:2ndcorrs}
\end{eqnarray}
The last line gives the special value of these correlations when
$r = 0.$

The matrix $\mathbf{K}_{\mathrm{c}}$ is the $4\times 4$ lower
right submatrix of $\bSigma_{\mathrm{c}},$ and has determinant
\begin{equation}
   \det \mathbf{K}_{\mathrm{c}} = (F_0^2 - H^2)(F_0^2 - F^2).
   \label{eq:critdetk}
\end{equation}
The pair of differential jacobian operators are, from
(\ref{eq:hessdef}),
\begin{equation}
   \mathcal{J}_{\nabla A} \mathcal{J}_{\nabla B}
   = \nabla_1 \nabla_2 \nabla_3 \nabla_4 + \nabla_5^2 \nabla_6^2
   -\nabla_1 \nabla_2 \nabla_6^2 - \nabla_3 \nabla_4 \nabla_5^2.
   \label{eq:jdifcrit}
\end{equation}

Using (\ref{eq:dexp}), the result of these operators acting is on
$\exp(-\bi{t}\cdot\bXi\cdot\bi{t}/2)$ and setting $\bi{t} =
\mathbf{0}$ (c.f. (\ref{eq:gabfin}),(\ref{eq:ddef})) is
\begin{equation}
\fl   D_{\mathrm{c}}
   = \Xi_{14} \Xi_{23} + \Xi_{13} \Xi_{24} - 2\Xi_{16} \Xi_{26} + \Xi_{12} \Xi_{34}
    - 2 \Xi_{35} \Xi_{45} - \Xi_{34} \Xi_{55} + 2 \Xi_{56}^2
    - \Xi_{12} \Xi_{66} + \Xi_{55} \Xi_{66},
    \label{eq:critdxexp}
\end{equation}
where the necessary entries of $\bXi_{\mathrm{c}},$ are found
using Jacobi's determinant theorem in (\ref{eq:xiij}); as an
example,
\begin{equation}
   \Xi_{24} = M_0 - F I^2/(F_0^2 - F^2).
   \label{eq:critxiexp}
\end{equation}

The topological charge correlation function $g_{\rm{c}}(r)$ for
critical points is obtained by substituting (\ref{eq:critdetk}),
(\ref{eq:critdxexp}) (with all terms like (\ref{eq:critxiexp})
found using (\ref{eq:xiij}) into (\ref{eq:gabfin})). This
expression is complicated and not very illuminating, and is not
given here. Upon substituting (\ref{eq:1stcorrs}),
(\ref{eq:1stcorr0}), (\ref{eq:2ndcorrs}) in, one finds that
$g_{\mathrm{c}}$ can be written as a perfect derivative (c.f.
(\ref{eq:vecgab})),
\begin{equation}
   g_{\mathrm{c}}(r) = \frac{1}{4 \pi^2 d_{\mathrm{c}}^2 r}
   \frac{\rmd h_{\mathrm{c}}}{\rmd r},
   \label{eq:critgab}
\end{equation}
where
\begin{eqnarray}
\fl   h_{\mathrm{c}}(r)  = \frac{(C'' - C'/r)}{r \sqrt{(C_0''^2 -
   C''^2)(C_0''^2 - C'^2/r^2)}} \left[ \frac{C^{(3)} (3C_0''^2 -
   2C''^2 - C'' C'/r)}{C_0''^2 - C''^2} \right. \nonumber \\
   \lo \quad \left. - \frac{(C'' -
   C'/r)(3C_0''^2 - 2C'^2/r^2-C'' C'/r)}{r(C_0''^2 - C'^2/r^2)}
   \right] \nonumber \\
   \lo = \frac{1}{r (C'' - C'/r)} \frac{\rmd}{\rmd r} \frac{(C'' - C'/r)^3}{\sqrt{(C_0''^2 -
   C''^2)(C_0''^2 - C'^2/r^2)}}.
   \label{eq:hcrit}
\end{eqnarray}
This process of finding $h_{\mathrm{c}}$ is long and tedious, and
details are omitted here. It is easy to see that
$h_{\mathrm{c}} \to 0$ as $r \to \infty;$ it is straightforward,
by Taylor expanding derivatives of $C,$ to show that
\begin{equation}
   h_{\mathrm{c}}(0) = \frac{4 C^{(4)}_0}{3 \sqrt{3} C''_0} =
   -2\pi d_{\mathrm{c}}.
   \label{eq:hc0}
\end{equation}

\begin{figure}
\begin{center}
\includegraphics*[width=12cm]{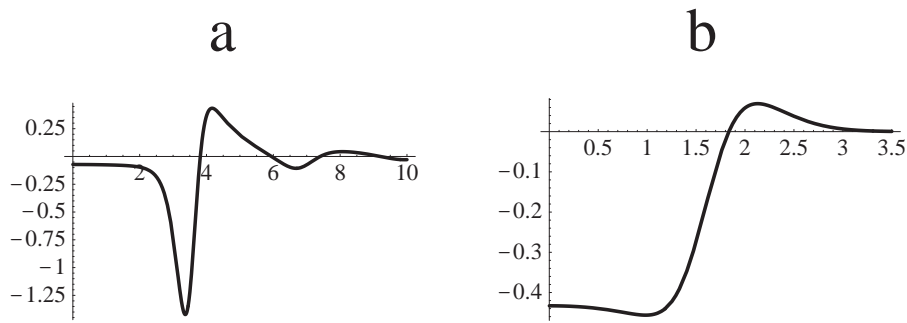}
\end{center}
    \caption{The critical point charge correlation function $g_{\mathbf{c}}(r),$
    plotted for two choices of $C(r):$ (a) $C(r) = J_0(r);$ (b) $C(r) = \exp(-r^2/2).$}
    \label{fig:gc}
 \end{figure}

The critical point topological charge correlation function for
two particular field correlation functions is shown in figure
\ref{fig:gc}. As with the 2-dimensional vector case plotted in
figure \ref{fig:g2}, the properties of the correlation function
(oscillatory, exponential decay, etc) are similar to that of the
underlying field correlation function $C(r),$ on which the
correlation function depends. Features of interest in these plots include the sharp initial minimum of (a), and the fact that the monotonic decay in (b) is from above, not below (in contrast to its counterpart in figure \ref{fig:g2}. Nevertheless,
the form of $h_{\mathrm{c}}$ is significantly more complicated
than $h_n,$ especially when $n = 2.$

\section{Umbilic points}\label{sec:umbilic}

Less well-known than critical points, umbilic points are geometric
point singularity features associated with the second derivative
of $f$ - namely where hessian matrix of second derivatives
$\partial_{ij} f$ is degenerate
(Berry and Hannay 1977, Porteous 2001, Hilbert and Cohn-Vossen 1952).
Geometrically, the principal axes of gaussian
curvature are not defined at these points. 
The eigenvalues of the hessian coincide when $f_{,xx} = f_{,yy},
\, f_{,xy} = 0.$ Umbilic points are therefore zeros of the
2-dimensional vector field
\begin{equation}
   \bi{v}_{\mathrm{u}} = \left((f_{,xx} - f_{,yy})/2, f_{,xy}\right).
   \label{eq:vumb}
\end{equation}
The factor of half in the first term ensures that $\bi{v}_{\mathrm{u}}$ is statistically rotation invariant.

An umbilic point has an index, determined geometrically by the
sense of rotation of the principal axes of curvature around the
umbilic point, and the index is generically $\pm 1/2$
(Berry and Hannay 1977); only the sign of the index is of interest here, and
this is determined by the appropriate jacobian
$\mathcal{J}_{\mathrm{u}}$ on $\bi{v}_{\mathrm{u}}$,
\begin{equation}
   2\mathcal{J}_{\mathrm{u}} = f_{,xxx} f_{,xyy} + f_{,yyy} f_{,xxy} -
   f_{,xyy}^2 - f_{,xxy}^2,
   \label{eq:jacumb}
\end{equation}
depending on the third partial derivatives of $f.$ The calculation
of the topological charge correlation function for umbilic points
can proceed according to the scheme of section \ref{sec:scheme},
in a similar way to the corresponding calculation for critical
points.

Umbilic points for isotropic random functions was considered by
Berry and Hannay; they found the density $d_{\mathrm{u}}$ to be (Berry and Hannay 1977 equation (34)) 
\begin{equation}
   d_{\mathrm{u}} = \left| \frac{3 C^{(6)}_0}{10\pi C^{(4)}_0} \right|,
   \label{eq:dumb}
\end{equation}
($C^{(6)}_0$ is the sixth derivative of $C$ at 0) and that the average index $q$ is zero (the separate densities are
$0.5d_{\mathrm{u}}$ for stars, $0.05279d_{\mathrm{u}}$ for
monstars, and $0.44721 d_{\mathrm{u}}$ for lemons). In the present
work, the distinction between monstars and lemons, which both have
positive index, is not used.

The ordering of the vector of gaussian random functions
$\bi{u}_{\mathrm{u}}$ is chosen
\begin{eqnarray}
   \bi{u}_{\mathrm{u}} =
   (f_{A,xxx}, f_{A,xyy}, f_{B,xxx}, f_{B,xyy}, f_{A,xxy}, f_{A,yyy},
   f_{B,xxy}, f_{B,yyy},\nonumber \\
   \qquad \qquad (f_{A,xx}-f_{A,yy})/2, (f_{B,xx}-f_{B,yy})/2, f_{A,xy}, f_{B,xy}).
   \label{eq:biguumb}
\end{eqnarray}
The correlation matrix (\ref{eq:cormatdef}) is
\begin{equation}
\fl   \bSigma_{\mathrm{u}} = \left( \begin{array}{cccccccccccc}
 S_0 &  T_0 &    S &    T &    0 &    0 &    0 &    0 &    0 &   -X & 0 & 0 \\
 T_0 &  T_0 &    T &    U &    0 &    0 &    0 &    0 &    0 &   -Y & 0 & 0 \\
   S &    T &  S_0 &  T_0 &    0 &    0 &    0 &    0 &    X &    0 & 0 & 0 \\
   T &    U &  T_0 &  T_0 &    0 &    0 &    0 &    0 &    Y &    0 & 0 & 0 \\
   0 &    0 &    0 &    0 &  T_0 &  T_0 &    T &    U &    0 &    0 & 0 & -Q \\
   0 &    0 &    0 &    0 &  T_0 &  S_0 &    U &    V &    0 &    0 & 0 & -R \\
   0 &    0 &    0 &    0 &    T &    U &  T_0 &  T_0 &    0 &    0 & Q & 0 \\
   0 &    0 &    0 &    0 &    U &    V &  T_0 &  S_0 &    0 &    0 & R & 0 \\
   0 &    0 &    X &    Y &    0 &    0 &    0 &    0 &  L_0 &    W & 0 & 0 \\
  -X &   -Y &    0 &    0 &    0 &    0 &    0 &    0 &    W &  L_0 & 0 & 0 \\
   0 &    0 &    0 &    0 &    0 &    0 &    Q &    R &    0 &    0 &L_0& L \\
   0 &    0 &    0 &    0 &   -Q &   -R &    0 &    0 &    0 &    0 & L&L_0 \\
   \end{array} \right),
   \label{eq:cormatumb}
\end{equation}
where $W \equiv (M+N-2L)/4, X \equiv (P-Q)/2, Y \equiv (Q-R)/2,$ and
\begin{eqnarray}
\fl   P & \equiv -\left[\partial_x^5 C\right]_{x=r,y=0} & = - C^{(5)}, \nonumber \\
\fl   Q & \equiv -\left[\partial_x^3 \partial_y^2 C\right]_{x=r,y=0} & = -(r^3 C^{(4)}-3r^2 C^{(3)}+ 6r C'' - 6 C')/r^4, \nonumber \\
\fl   R & \equiv -\left[\partial_x \partial_y^4 C\right]_{x=r,y=0} & = -3(r^2 C^{(3)} - 3 r C'' + 3 C')/r^4, \nonumber \\
\fl   S & \equiv-\left[\partial_x^6 C\right]_{x=r,y=0} & = - C^{(6)}, \nonumber \\
\fl   T & \equiv -\left[\partial_x^4 \partial_y^2 C\right]_{x=r,y=0} & =  -(r^4 C^{(5)} - 4 r^3 C^{(4)}
   + 12 r^2 C^{(3)} - 24 r C'' + 24 C')/r^5, \nonumber \\
\fl   U & \equiv  -\left[\partial_x^2 \partial_y^4 C\right]_{x=r,y=0} & = -3(r^3 C^{(4)} - 5 r^2 C^{(3)} + 12 r C'' - 12 C')/r^5, \nonumber \\
\fl   V & \equiv -\left[\partial_y^6 C\right]_{x=r,y=0} & = -15(r^2 C^{(3)} - 3 r C'' + 3 C')/r^5, \nonumber \\
\fl   P_0 & = Q_0 = R_0 = 0, & S_0 = 5T_0 = 5U_0 = V_0 = -C^{(6)}_0.
   \label{eq:3rdcorrs}
\end{eqnarray}

\begin{figure}
\begin{center}
\includegraphics*[width=12cm]{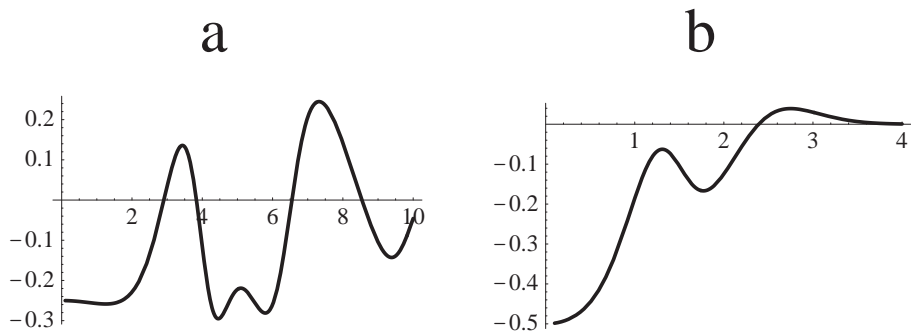}
\end{center}
    \caption{The umbilic charge correlation function $g_{\mathbf{u}}(r),$
    plotted for two choices of $C(r):$ (a) $C(r) = J_0(r);$ (b) $C(r) = \exp(-r^2/2).$}
    \label{fig:gu}
\end{figure}

The matrix $\mathbf{K}_{\mathrm{u}},$ defined using (\ref{eq:sigkxidef}), has
determinant
\begin{equation}
   \det \mathbf{K}_{\mathrm{u}} = (L_0^2 - (M + N -2L)^2/16)(L_0^2
   - L^2).
   \label{eq:detkumb}
\end{equation}
The result of the jacobian derivative operators (\ref{eq:ddef})
gives
\begin{eqnarray}
\fl   D_{\mathrm{u}} =
   \Xi_{12}^2 + \Xi_{14}^2  + \Xi_{22}^2 + 2 \Xi_{24}^2 - 2\Xi_{12} \Xi_{22}
   + \Xi_{13} \Xi_{24} - 4 \Xi_{14} \Xi_{24}
   - 2 \Xi_{12} \Xi_{55} + 2\Xi_{12}\Xi_{56}
   \nonumber \\ - 2\Xi_{22} \Xi_{56}
   + \Xi_{55}^2 + \Xi_{56}^2 + 2\Xi_{57}^2 + \Xi_{58}^2
   - 2 \Xi_{55} \Xi_{56}
   - 4 \Xi_{57} \Xi_{58} + \Xi_{57} \Xi_{68}.
   \label{eq:umbdxexp}
\end{eqnarray}
The necessary entries of $\bXi_{\mathrm{u}}$ are found using
(\ref{eq:xiij}). The resulting expression for $D_{\mathrm{u}},$
and therefore for $g_{\mathrm{u}},$ is very complicated, but may be
reduced to the following form:
\begin{equation}
   g_{\mathrm{u}}(r) = \frac{1}{4\pi^2 d_{\mathrm{u}}^2 r} \frac{\rmd h_{\mathrm{u}}}{\rmd r},
   \label{eq:gumb}
\end{equation}
where
\begin{equation}
   h_{\mathrm{u}}(r) = \frac{\rmd}{\rmd r}\left( \frac{r (Q - R)^2}
   {4\sqrt{\det \mathbf{K}_{\mathrm{u}}}} \right) +
   \frac{Q (P + R - 2Q)}{4\sqrt{\det \mathbf{K}_\mathrm{u}}}.
   \label{eq:humb}
\end{equation}
It can be shown that $h_{\mathrm{u}}(0) = -2\pi d_{\mathrm{u}},$ and $h \to 0$ as $r\to \infty.$ $g_{\mathrm{u}}(r)$ is plotted in figure \ref{fig:gu} for $C(r) =
J_0(r)$ and $\exp(-r^2/2).$ The most obvious feature of these two
plots, compared to figures \ref{fig:g2}, \ref{fig:gc}, is that
they have a negative maximum, a property that seems to be general
for $g_{\mathrm{u}},$ although this has not been proved. It is
unclear what the physical significance of this kink should be;
mathematically, it probably arises from interference between the
two summands in $h_{\mathrm{u}}(r)$ in (\ref{eq:humb}).

\section{Topological charge screening}\label{sec:screening}

Three particular charge correlation functions have been derived
exactly (equations (\ref{eq:vecgab}), (\ref{eq:hndef}), (\ref{eq:critgab}), (\ref{eq:hcrit}), (\ref{eq:gumb}),(\ref{eq:humb})). In each case, the charge correlation function has the form
\begin{equation}
   g(r) = \frac{(n-1)!}{(2\pi)^n d^2 r^{n-1}} \frac{\rmd h(r)}{\rmd r},
   \label{eq:ggen}
\end{equation}
where $n$ is the number of dimensions, $d$ is the density of zeros and
$h(r)$ is a function such that
\begin{equation}
   h(0) = \frac{(2\pi)^n d}{\sigma_{n-1} (n-1)!}, \qquad h\to \infty \mathrm{\, as \,}
   r \to \infty.
   \label{eq:hgen}
\end{equation}
The total charge density around a given (say positive) topological charge is therefore
\begin{eqnarray}
   d \int \rmd^n \bi{r} \, g(r) \nonumber & = & \frac{(n-1)!}{(2\pi)^n d}
   \sigma_{n-1} \int_0^{\infty} \rmd r \, \frac{\rmd h(r)}{\rmd r}
   \nonumber \\
   & = & \frac{(n-1)! \sigma_{n-1}}{(2\pi)^n d^2} (h(\infty) - h(0)) \nonumber \\
   & = & -1,
   \label{eq:screening1}
\end{eqnarray}
where the hypersphere area $\sigma_{n-1}$ appears in the first
line from integration in polar coordinates. It implies that the
distribution of topological charges is such that every topological
charge tends to be surrounded by charges of the opposite sign,
such that the topological charge is `screened' at large distances.
This fact was noticed in the random vector case by
Halperin (1981) and Liu and Mazenko (1992) (although the general zero density
$d_n$ had not been determined explicitly) and is independent of
the field correlation function $C.$ The derivation here shows
that this is a more general phenomenon, possibly a universal
feature of topological charge correlations for gaussian random
fields. It should be noted that (\ref{eq:screening1}) is not
necessarily satisfied for an arbitrary distribution of signed
points; for instance, $g(r) = 0$ always for Poisson points, for
which there is no screening. $\delta$-function correlations at the origin are ignored in the following.

An analogy may be drawn from the theory of ionic liquids
(Hansen and McDonald 1986); in a fluid or plasma, consisting of two species identical
except for opposite (Coulomb) charges, the following
Stillinger-Lovett sum rules (Stillinger and Lovett 1968a,b)
 are found to hold:
\begin{eqnarray}
   d \int \rmd^n \bi{r} \, g(r) & = & -1, \label{eq:slsumrule1} \\
   d \int \rmd^n \bi{r} \, r^2 g(r) & = & -a_n \Lambda^2.
   \label{eq:slsumrule2}
\end{eqnarray}
Here, $g$ is the charge-charge correlation function, $\Lambda$ is a characteristic screening length dependent on
temperature and $n,$ and $a_n$ is a constant dependent on
dimensionality. These rules are discussed for $n = 3$ by Hansen and McDonald (1986), Stillinger and Lovett 1968a,b, and $n = 2$ by Jancovici 1987, Jancovici \etal 1994.

The screening relation (\ref{eq:screening1}) is equivalent to the first Stillinger-Lovett
sum rule (\ref{eq:slsumrule1}), which is derived using the statistical
mechanics of pairwise, Coulomb interacting fluids. It is unclear
whether the fact that topological and coulombic charges screen in
the same way is coincidence, or evidence of some deeper connection
between the two statistical theories.

It is natural to ask whether the topological charge correlation
functions satisfy the second sum rule, which (upon integrating the
left hand side of (\ref{eq:slsumrule2}) by parts), depends on the
integral of $r h.$ For $n=2,$ it was
found (Berry and Dennis 2000) that for certain choices of $C,$ this integral may
diverge. The slowest decay comes when $C = J_0,$ and by
(\ref{eq:hndef}),
\begin{equation}
   h_2(r) \sim_{r\to \infty} \cos(r+\pi/4)^2/r \qquad \mathrm{when} \quad C = J_0,
   \label{eq:h2asymp}
\end{equation}
giving a logarithmic divergence for the second moment. For
comparison, the critical and umbilic functions
$h_{\mathrm{c}},h_{\mathrm{u}},$ (equations (\ref{eq:hcrit}),
(\ref{eq:humb}) respectively) both give, for the same choice of $C,$
\begin{equation}
   h_{\mathrm{c}}(r),h_{\mathrm{u}}(r)  \sim_{r\to \infty} \cos(2r)/r^2 \qquad \mathrm{when}
   \quad C =
   J_0,
   \label{eq:hcuasymp}
\end{equation}
implying that the second sum rule is satisfied generally for
critical and umbilic points, although, as for the cases of random
vectors where the integral converges, the screening length
$\Lambda,$ defined in analogy to (\ref{eq:slsumrule2}), depends on the choice of $C.$ For random vectors with $n>2,$ the second moment of $g$ always converges, because of the higher power of $C'$ appearing in (\ref{eq:hndef}) (also, the decay of $C$ will be faster, as discussed at the end of section \ref{sec:fields}).

Comparison may be drawn to the electrostatic analogy in
random matrix theory, particularly in the case of the
so-called Ginibre ensemble of $N \times N$ matrices whose entries
are independent, identically distributed circular gaussian random
variables (Ginibre 1965). The eigenvalues of these
matrices are found to have exactly the same statistical behaviour
as a 1-component 2-dimensional Coulomb gas of $N$ charges 
in a harmonic oscillator potential, and the 2-point
density correlation function screens against a uniform background
(i.e. $d \int \rmd^2 z \, (g_{\mathrm{Gin}} -1) = -1$), and have
finite second moment. Certain random polynomial analogues have
zeros that can also be expressed as 2-dimensional Coulomb gases
with additional interactions (Hannay 1998, Forrester and Honner 1999). The eigenvalues of random
matrices (which may be expressed as the zeros of the
characteristic polynomial) and zeros of random polynomials are all
of the same sign, since they are zeros of complex analytic
functions, and the density correlation functions of zeros are unique
(there is no analogue of $C$).

There may be a danger in taking the analogy with fluids too far; for instance, the oscillations of the functions in figures \ref{fig:g2},\ref{fig:gc},\ref{fig:gu}a are reminiscent of those of charged fluids (e.g. Hansen and McDonald 1986). However, the physical causes for these oscillations are very different. In fluids, the oscillations usually arise from packing considerations (the ions themselves are of finite size, fixing the lengthscale, although in plasmas they are usually represented as point charges (Baus and Hansen 1980)). Topological charges, on the other hand, are points, and the oscillations in these figures originate from the oscillations in the underlying field correlation function $C(r),$ which is $J_0$ in this case. Although Halperin (1981) discusses the similarity between the short-range behaviour of the 2-dimensional vector correlation function (\ref{eq:vecgab}), (\ref{eq:hndef}) and Kosterlitz-Thouless theory, the present situation is more general, both in that the (possibly long-range) screening is exact, and that the results hold for any reasonable field correlation function $C(r).$

\section{Discussion}

Using a general scheme for calculating topological charge
correlation functions, three particular correlation functions were
found explicitly, and were found always to satisfy a screening
relation (\ref{eq:screening1}).

The scheme of section \ref{sec:scheme} used to calculate the
charge correlation functions is very general, and can be
generalised to calculate the charge-charge correlation function
between two different sets of topological charges - for instance
with a critical point at $A,$ and an umbilic point at $B.$
Although not done so here, the scheme may be applied to
anisotropic fields.

It is rather more difficult to calculate the density correlation
function for topological charges (the analogue of (\ref{eq:gdef})
where the moduli of the jacobians are taken). It was calculated
for 1-dimensional vectors (i.e. random functions in 1 dimension)
by Rice (1954), and for 2-dimensional vectors (realised as complex
scalars) by Berry and Dennis (2000), Saichev \etal (2001). However, it has not been
possible to generalise such methods to the case of critical
points. Also, numerical evidence (Freund and Wilkinson 1998) suggests that
the correlation function of extrema signed by the sign of their
laplacian ($+1$ for minima, $-1$ for maxima) also satisfies the
screening relation (\ref{eq:slsumrule1}). All such functions would
be needed to calculate the partial correlation functions between
the species (e.g. maxima with maxima, maxima with saddles, maxima
with minima, etc), which would give a complete statistical
picture.

The presence of boundaries in the random function will affect the
statistical properties of topological charges (e.g.
Berry (2002) for nodal points in the plane), and it is
possible that there may be some further analogy with the physics
of interfaces of Coulomb fluids. The scheme of section \ref{sec:scheme} ought to be adaptable to calculate charge correlation functions in this case.

Only zeros of fields linear in gaussian random functions have been
considered here, although the density of others may be calculated,
for instance, in addition to nodal points, a 2-dimensional
gaussian random complex scalar has critical points of its modulus
squared (Weinrib and Halperin 1982) and its argument
(Dennis 2001a). The scheme employed here cannot be used to
calculate correlation functions for these, although numerical
evidence (I Freund, personal communication) suggests that the
critical points of argument (together with the nodal points) do
screen, therefore adding to the cases shown here. It is tempting to conjecture that
topological charge screening may be a universal phenomenon in
gaussian random fields.

\ack

I would like to thank Michael Berry and Robert Evans for useful discussions, John Hannay for discussions leading to the arguments in Appendix B, and Isaac Freund for correspondence. This research was supported by the Leverhulme Trust.

\appendix

\section{Jacobi's determinant theorem}\label{sec:jacobi}

Let $\mathbf{A}$ be a square matrix. The {\itshape
minor}  $\mathcal{M}^{i_1
\dots i_k}_{j_1 \dots j_k}(\mathbf{A})$ is the determinant of the
$k \times k$ submatrix of $\mathbf{A}$ with rows $i_1, \dots, i_k$
and columns $j_1, \dots, j_k.$ $\overline{\mathcal{M}}^{i_1, \dots,
i_k}_{j_1, \dots, j_k}(\mathbf{A})$ shall be used to denote the
complementary minor, that is, the minor of the submatrix of
$\mathbf{A}$ with rows $i_1, \dots, i_k$ and columns $j_1, \dots,
j_n$ excluded. Then Jacobi's determinant theorem (Jeffreys and Jeffreys 1956, page 135) states 
\begin{equation}
   \det\mathbf{A} \, \mathcal{M}^{i_1, \dots, i_k}_{j_1, \dots,
   j_k}(\mathbf{A}^{-1})= (-1)^{i_1+\dots+i_k+j_1+\dots+j_k}
   \overline{\mathcal{M}}^{i_1, \dots, i_k}_{j_1, \dots,
   j_k}(\mathbf{A}).
   \label{eq:jacobith}
\end{equation}
Applying this to $\bSigma$ in (\ref{eq:sigkxidef}), and choosing
$\bXi^{-1}$ as the submatrix whose determinant is the minor of $\bSigma^{-1},$
\begin{eqnarray}
   \det\bSigma \, \det\bXi^{-1} & = & \det\bSigma \, \mathcal{M}^{1,
   \dots, 2m}_{1, \dots, 2m}(\bSigma^{-1}) \nonumber \\
   & = & (-1)^{1+\dots+2m+1+\dots+2m} \mathcal{M}^{2m+1, \dots, 2(m+n)}_{2m+1, \dots,
   2(m+n)}(\bSigma) \nonumber \\
   & = & \det\mathbf{K}, \label{eq:jacobiproof}
\end{eqnarray}
from which (\ref{eq:jacobi}) follows directly.

Jacobi's theorem can also be used to find the elements $\Xi_{ij}$
of the inverse reduced inverse matrix $\bXi$ in
(\ref{eq:sigkxidef}), needed for (\ref{eq:dexp}). In this case,
(\ref{eq:jacobith}) is applied twice, once on the matrix
$\bXi^{-1},$ and once on $\bSigma.$ Therefore
\begin{eqnarray}
   \Xi_{ij} & = & \mathcal{M}^i_j(\bXi) \nonumber \\
   & = & (-1)^{i+j} \overline{\mathcal{M}}^i_j(\bXi^{-1})/\det \bXi^{-1} \nonumber \\
   & = & (-1)^{i+j}\det{\bXi} \, \mathcal{M}^{1, \dots, i-1, i+1, \dots, 2m}_{1, \dots,
   j-1, j+1, \dots, 2m}(\bSigma^{-1}) \nonumber \\
   & = & \det{\bXi} \, \mathcal{M}^{i,2m+1,\dots, 2m+2n}_{j,2m+1,\dots,
   2m+2n}(\bSigma)\nonumber \\
   & = & \mathcal{M}^{i,2m+1,\dots, 2m+2n)}_{j,2m+1,\dots,
   2m+2n}(\bSigma) /\det \mathbf{K}, \nonumber \\
   & = & \det \mathbf{K}^{-1} \det \left( \begin{array}{cc} \Sigma_{ij} & \bullet
   \\ \bullet & \mathbf{K}
   \end{array}\right)
\label{eq:xiij}
\end{eqnarray}
where in the last line $\bullet$ represents the terms
$\Sigma_{i,k}, \Sigma_{k,j}$ where $k = 2m+1, \dots, 2m+2n.$ Thus
the $\Xi_{ij}$ appearing in the expression for $D$ in (\ref{eq:dexp}), is the
determinant of the $(2n+1)\times (2n+1)$ submatrix comprised of
the $i$th row and $j$th column of $\bSigma,$ and the submatrix
$\mathbf{K}.$

\section{Calculation of the density of vector zeros
(\ref{eq:dveccalc})} \label{sec:dveccalc}

In order to integrate (\ref{eq:dveccalc}), the following must be integrated
\begin{equation}
   \mathcal{V} = \int\rmd^{n^2} v_{i,j} \,
   |\det v_{i,j} | \exp\left(-\frac{1}{2}\sum_{i,j=1}^n v_{i,j}^2\right).
   \label{eq:dveccalc1}
\end{equation}
This is, mathematically, the average hypervolume of an $n$-dimensional
parallelepiped specified by gaussian random vectors $\bi{w}_1 =
(v_{1,1}, \dots v_{1,n}), \dots, \bi{w}_n = (v_{n,1}, \dots,
v_{n,n}).$ These gaussian random vectors are identically
distributed isotropically in $n$-dimensional space. The
hypervolume is nonzero exactly when the set of $n$ vectors is
linearly independent.

This hypervolume may be found explicitly in a manner
reminiscent of the Gram-Schmidt orthogonalization procedure for
vectors. Geometrically,
\begin{eqnarray}
\fl   \rm{volume \, of \, parallelepiped} = \rm{length \, of \,}
\bi{w}_1 \times
   \rm{\,length \, of \,} \bi{w}_2 \rm{\, orthogonal \, to \,} \bi{w}_1 \times \dots
    \nonumber \\ \times
   \rm{\, length \, of \,} \bi{w}_n {\, \rm orthogonal  \, to \,}
    \mathrm{span}\{\bi{w}_1,\dots,\bi{w}_{n-1}\}.
   \label{eq:parallelgeom}
\end{eqnarray}
A given factor in this product is therefore the average length of
the gaussian random vector $\bi{w}_j$ in the orthogonal complement
of a $(j-1)$-dimensional subspace $\mathrm{span}\{\bi{w}_1, \dots, \bi{w}_{j-1}\}.$

Now, since the vector $\bi{w}_j$ is isotropic, it may be
represented identically in any choice of orthonormal basis of
$n$-dimensional space; in particular, its first $j-1$ components
$v_{j,1}, \dots, v_{j,j-1}$ may be chosen to be in
$\mathrm{span}\{\bi{w}_1, \dots, \bi{w}_{j-1}\}$ (as in the
Gram-Schmidt procedure). The total contribution of $\bi{w}_j$ to
the integral in (\ref{eq:dveccalc}) involves the average length of
the vector made up of the other components $v_{j,j}, \dots,
v_{j,n}.$ Where $k = n - j -1,$ this is
\begin{eqnarray}
   \int \rmd^n \bi{w}_j \, \sqrt{v_{j,1}^2 +\dots +v_{j,n}^2} \exp\left(-\frac{1}{2}
   \sum_{i = 1}^n v_{j,i}^2\right) \nonumber \\
   = \left[\int_{-\infty}^{\infty} \rmd v_{j,1} \,
   \exp(-v_{j,1}^2/2)\right]^{j-1} \int_{\mathcal{S}_{k-1}}
   \rmd^{k-1}\Omega_{k-1} \int_0^{\infty} \rmd \rho \, \rho^k
   \exp(-\rho^2/2)
   \label{eq:hyperint}
\end{eqnarray}
where, in the second line, the first $m-1$ components have been
pulled out as trivial gaussians, integrating to $(2\pi)^{(j-1)/2}
= (2\pi)^{(n-k)/2};$ the remaining $k$ integrals are the average
length of a gaussian random vector in $k$-dimensional space. This
integral has been converted to polar coordinates,  with
$\rmd\omega_{k-1}$ the solid angle infinitesimal on the unit
$(k-1)$-sphere $\mathcal{S}_{k-1},$ and $\rho$ is the radius. It
is well known that the surface area $\sigma_{k-1}$ of the unit
$(k-1)$-hypersphere is
\begin{equation}
   \sigma_{k-1} = \int \rmd^{k-1}\Omega_{k-1} =
   \frac{2\pi^{k/2}}{(\frac{k-2}{2})!}.
   \label{eq:spherearea}
\end{equation}
The $\rho$ integral in (\ref{eq:hyperint}) is
$2^{(k-1)/2}((k-1)/2)!.$ Therefore, the numerical part of
(\ref{eq:dveccalc}) is $1/(2\pi)^{n(n+1)/2}$ times the product in
(\ref{eq:parallelgeom}), with each term in the product, now
labelled by $k,$ given by the expression (\ref{eq:hyperint}).
Therefore
\begin{eqnarray}
   \mathcal{V} & =& \frac{1}{(2\pi)^{n(n+1)/2}} \prod_{k=1}^{n} (2\pi)^{(n-k)/2}
   \times \frac{2\pi^{k/2}}{(\frac{k-2}{2})!}\times
   2^{(k-1)/2}\left(\frac{k-1}{2}\right)! \nonumber \\
   & = & \frac{1}{(2\pi)^{n(n+1)/2}} \prod_{k=1}^n (2\pi)^{(n+1)/2}
   \pi^{n/2} \frac{(\frac{k-1}{2})!}{(\frac{k-2}{2})!} \nonumber
   \\
   & = & \frac{(\frac{n-1}{2})!}{\pi^{(n+1)/2}}.
   \label{eq:prodcalc}
\end{eqnarray}
This value agrees with that stated by Halperin (1981), Liu and Mazenko (1992) of $1/\pi$ ($n=1$), $1/2\pi$ ($n
= 2$) and $1/\pi^2$ ($n = 3$).

\References

\item[]
Adler R J 1981
\newblock {\em The geometry of random fields}
\newblock (Wiley)

\item[]
Baus M and Hansen J-P 1980
\newblock Statistical mechanics of simple Coulomb systems
\newblock {\em Phys.Rep.} {\bf 59} 1--94

\item[]
Berry M V 1977
\newblock Regular and irregular semiclassical wavefunctions
\newblock {\em J.Phys.A:Math.Gen.} {\bf 10} 2083--91

\item[]
\dash 1978
\newblock Disruption of wavefronts: statistics of dislocations in incoherent
  gaussian random waves
\newblock {\em J.Phys.A:Math.Gen.} {\bf 11} 27--37

\item[]
\dash 2002
\newblock Statistics of nodal lines and points in quantum billiards: perimeter corrections, fluctuations, curvature
\newblock {\em J.Phys.A:Math.Gen.} {\bf 35} 3025--38

\item[]
Berry M V and Hannay J H 1977
\newblock Umbilic points on a gaussian random surface
\newblock {\em J.Phys.A:Math.Gen.} {\bf 10} 1809--21

\item[]
Berry M V and Dennis M R 2000
\newblock Phase singularities in isotropic random waves
\newblock {\em Proc.R.Soc.Lond.A.} {\bf 456} 2059--79
\newblock (errata {\bf 456} 3059).

\item[]
Dennis M R 2001a
\newblock Phase critical point densities in planar isotropic random waves
\newblock {\em J.Phys.A:Math.Gen.} {\bf 34} L297--L303

\item[]
\dash 2001b
\newblock {\em Topological singularities in wave fields}
\newblock Ph.D. thesis, Bristol University

\item[]
\dash 2002
\newblock Polarization singularities in paraxial vector fields: morphology and
  statistics
\newblock {\em Opt.Commun.} {\bf 213} 201--21

\item[]
Feller W 1950
\newblock {\em An introduction to probability theory and its applications},
  volume~I.
\newblock (Wiley, New York)

\item[]
Foltin G 2003a
\newblock Signed zeros of gaussian vector fields - density, correlation
  functions and curvature
\newblock {\em J.Phys.A:Math.Gen.} {\bf 36} 1729--41

\item[]
\dash 2003b
\newblock The distribution of extremal points of Gaussian scalar fields.
\newblock {\em J.Phys.A:Math.Gen.} {\bf 36} 4561--80

\item[]
Forrester P J and Honner G 1999
\newblock {Exact statistical properties of complex random polynomials}
\newblock {\em J.Phys.A:Math.Gen.} {\bf 32} 2961--81

\item[]
Freund I and Wilkinson M 1998 
\newblock Critical-point screening in random wave fields
\newblock {\em J.Opt.Soc.Am.A} {\bf 15} 2892--902

\item[]
Ginibre J 1965
\newblock Statistical ensembles of complex, quaternion and real matrices.
\newblock {\em J.Math.Phys.} {\bf 6} 440--49

\item[]
Halperin B I 1981
\newblock Statistical mechanics of topological defects.
\newblock in R~Balian, M~Kl{\'e}man, and J-P Poirier, eds, {\em Les
  Houches Session XXV - Physics of Defects} (North-Holland, Amsterdam)

\item[]
Hannay J H 1998
\newblock {The chaotic analytic function}
\newblock {\em J.Phys.A:Math.Gen.} {\bf 31} L755--61

\item[]
Hansen J-P and McDonald I R 1986
\newblock {\em Theory of simple liquids}
\newblock (Academic Press)

\item[]
Hilbert D and Cohn-Vossen S 1952
\newblock {\em Geometry and the Imagination}
\newblock (Chelsea Publishing)

\item[]
Jancovici B 1987
\newblock Charge correlations and sum rules in Coulomb systems. I.
\newblock in F J Rogers and H E Dewitt, eds, {\em Strongly Coupled Plasma Physics} (Plenum)

\item[]
Jancovici B, Manificat G and Pisani C 1994
\newblock Coulomb systems seen as critical systems: Finite-size effects in two dimensions
\newblock {\em J.Stat.Phys.} {\bf 78} 307--29

\item[]
Jeffreys H and Jeffreys B S 1956
\newblock {\em Methods of Mathematical Physics}
\newblock (Cambridge University Press)

\item[]
Liu F and Mazenko G F 1992
\newblock Defect-defect correlation in the dynamics of first-order phase
  transitions
\newblock {\em Phys.Rev.B} {\bf 46} 5963--71

\item[]
Longuet-Higgins M S 1957a
\newblock The statistical analysis of a random, moving surface
\newblock {\em Phil.Trans.R.Soc.A}, {\bf 249} 321--87

\item[]
\dash 1957b
\newblock Statistical properties of an isotropic random surface
\newblock {\em Phil.Trans.R.Soc.A}  {\bf 250} 157--74

\item[]
\dash 1958
\newblock The statistical distribution of the curvature of a random Gaussian
  surface
\newblock {\em Proc.Camb.Phil.Soc.} {\bf 54} 439--53

\item[]
Mermin N D 1979
\newblock The topological theory of defects in ordered media
\newblock {\em Rev.Mod.Phys.} {\bf 51} 591--648

\item[]
Milnor J W 1965
\newblock {\em Topology from the differentiable viewpoint}
\newblock (Virginia University Press)

\item[]
Nye J F and Berry M V 1974
\newblock Dislocations in wave trains
\newblock {\em Proc.R.Soc.Lond.A} {\bf 336} 165--90

\item[]
Porteous I R 2001
\newblock {\em Geometric differentiation: for the intelligence of curves and
  surfaces,}
\newblock 2nd ed (Cambridge University Press) 

\item[]
Rice S O
\newblock Mathematical analysis of random noise,
\newblock reprinted in  N Wax, ed, 1954 {\em Selected papers on noise and stochastic processes} (Dover, New York)

\item[]
Saichev A I, Berggren K-F, and Sadreev A F 2001
\newblock Distribution of nearest distances between nodal points for the Berry
  function in two dimensions.
\newblock {\em Phys.Rev.E}  {\bf 64} 036222

\item[]
Stillinger F H and Lovett R 1968a
\newblock Ion-pair theory of concentrated electrolytes. I. Basic concepts.
\newblock {\em J.Chem.Phys.} {\bf 48} 3858--68

\item[]
\dash 1968b
\newblock General restriction on the distribution of ions in electrolytes
\newblock {\em J.Chem.Phys.} {\bf 49} 1991--4

\item[]
Weinrib A and Halperin B I 1982
\newblock Distribution of maxima, minima, and saddle points of the intensity of
  laser speckle patterns
\newblock {\em Phys.Rev.B} {\bf 26} 1362--8

\endrefs

\end{document}